\documentclass[USenglish,twocolumn]{article}

\newcommand{\co}     {$^{13}$CO\,(1--0)}
\newcommand{\cvo}     {C$^{18}$O\,(1--0)}
\newcommand{\cs}     {CS\,(2--1)}
\newcommand{\kms}    {km\,s$^{-1}$}

\newcommand{\hii}    {H\,{\sc{ii}}}
\newcommand{\ncs}    {$N_{\rm CS}$}
\newcommand{\nmeth}  {$N_{\rm CH_3OH}/\Delta V$}
\newcommand{\tk}     {$T_{\rm k}$}

\newcommand{\abull}{Astrophys~Bull}
\newcommand{\arep}{Astron~Reports}

\usepackage[utf8]{inputenc}
\usepackage[big]{dgruyter}
\usepackage{natbib}
\usepackage{upgreek}
\usepackage{amsmath,amsfonts,amssymb,pxfonts,eulervm,xspace}
\usepackage{graphicx}
\usepackage[title]{appendix}

\def\jref@jnl#1{{\rm#1\/}}
\def\actaa{\jref@jnl{Acta Astronomica}}
\def\aap{\jref@jnl{A\&A}}
\def\aapr{\jref@jnl{The Astronomy and Astrophysics Review}}
\def\aaps{\jref@jnl{Astronomy and Astrophysics Supplement Series}}
\def\aj{\jref@jnl{AJ}}
\def\apj{\jref@jnl{ApJ}}
\def\apjl{\jref@jnl{ApJL}}
\def\apjs{\jref@jnl{ApJS}}
\def\apss{\jref@jnl{Astrophysics and Space Science}}
\def\ao{\jref@jnl{Applied Optics}}
\def\araa{\jref@jnl{ARA\&A}}
\def\bain{\jref@jnl{BAN}}
\def\caa{\jref@jnl{Chinese Astronomy and Astrophysics}}
\def\cjaa{\jref@jnl{Chinese Journal of Astronomy and Astrophysics}}
\def\gca{\jref@jnl{Geochimica et Cosmochimica Acta}}
\def\jcp{\jref@jnl{Journal of Chemical Physics}}
\def\jqsrt{\jref@jnl{Journal of Quantitative Spectroscopy and Radiative Transfer}}
\def\mnras{\jref@jnl{MNRAS}}
\def\memras{\jref@jnl{Memoirs of the Royal Astronomical Society}}
\def\memsai{\jref@jnl{Memorie della Societa Astronomica Italiana}}
\def\na{\jref@jnl{New Astronomy}}
\def\nar{\jref@jnl{New Astronomy Reviews}}
\def\nat{\jref@jnl{Nature}}
\def\pasa{\jref@jnl{Publications of the Astronomical Society of Australia}}
\def\planss{\jref@jnl{Planetary and Space Science}}
\def\pasj{\jref@jnl{Publications of the Astronomical Society of Japan}}
\def\pasp{\jref@jnl{PASP}}
\def\physrep{\jref@jnl{Physics Reports}}
\def\rmxaa{\jref@jnl{Revista Mexicana de Astronomia y Astrofisica}}
\def\skytel{\jref@jnl{Sky and Telescope}}
\def\solphys{\jref@jnl{Solar Physics}}
\def\sovast{\jref@jnl{Soviet Astronomy}}
\def\ssr{\jref@jnl{Space Science Reviews}}
\def\zap{\jref@jnl{Zeitschrift fuer Astrophysik}}

\begin{document}

  \articletype{Research Article{\hfill}Open Access}

  \author*[1]{Maria S. Kirsanova}

  \author[2]{Svetlana V. Salii}

  \author[2]{Andrej M. Sobolev}

  \author[2]{Anders Olof Henrik Olofsson}

  \author[2]{Dmitry A. Ladeyschikov}

  \author[2]{Magnus Thomasson}

  \title{\huge Molecular gas in high-mass filament WB\,673}

  \runningtitle{Molecular gas in high-mass filament WB\,673}


\begin{abstract}
{We studied the distribution of dense gas in a filamentary molecular cloud containing several dense clumps. The center of the filament is given by the dense clump WB\,673. The clumps are high-mass and intermediate-mass star-forming regions. We observed \cs, \co, \cvo,\ and methanol lines at 96~GHz toward WB\,673 with the Onsala Space Observatory 20-m telescope. We found \cs\ emission in the inter-clump medium so the clumps are physically connected and the whole cloud is indeed a filament. Its total mass is $10^4$\,M$_{\odot}$ and mass-to-length ratio is 360\,M$_{\odot}$\,pc$^{-1}$ from \co\ data. Mass-to-length ratio for the dense gas is $3.4-34$\,M$_{\odot}$\,pc$^{-1}$ from \cs\ data. The PV-diagram of the filament is V-shaped. We estimated physical conditions in the molecular gas using methanol lines. Location of the filament on the sky between extended shells suggests that it could be a good example to test theoretical models of formation of the filaments via multiple compression of interstellar gas by supersonic waves.}
\end{abstract}

  \journalname{Open Astronomy}
\DOI{https://doi.org/10.1515/astro-2017-0020}
  \startpage{1}
  \received{Sep 14, 2017}
  \revised{..}
  \accepted{Oct 20, 2017}

  \journalyear{2017}
  \journalvolume{26}

\maketitle

{ \let\thempfn\relax
\footnotetext{\hspace{-1ex}{\Authfont\small \textbf{Corresponding Author: Maria S. Kirsanova}} {\Affilfont Institute of Astronomy, Russian Academy of Sciences, 48 Pyatnitskaya Str., Moscow, Russia, E-mail: kirsanova@inasan.ru}}
}

{ \let\thempfn\relax
\footnotetext{\hspace{-1ex}{\Authfont\small \textbf{Svetlana V. Salii, Andrej M. Sobolev, Dmitry A. Ladeyschikov}} {\Affilfont Ural Federal University, 19 Mira Str., Ekaterinburg, Russia}}
}

{ \let\thempfn\relax
\footnotetext{\hspace{-1ex}{\Authfont\small \textbf{Anders Olof Henrik Olofsson, Magnus Thomasson}} {\Affilfont Department of Space, Earth and Environment, Chalmers University of Technology, Onsala Space Observatory, SE-439 92 Onsala, Sweden}}
}

\section{Introduction}\label{sec:intro}

Molecular clouds often appear as elongated filamentary structures. Observations with the Herschel Space Observatory provide evidence that filamentary structure is a ubiquitous property of molecular clouds, see~\citet{Andre_2014}. Theoretical calculations, e.g.~\citet{Inutsuka_2015}, predict formation of filamentary molecular clouds after multiple compression of interstellar gas by supersonic waves. So the periphery of extended bubbles (\hii\ regions or supernova remnants) could be reliable places for the formation of the molecular filaments. We have thus initiated a study of giant molecular clouds which contain multiple bubbles to study the properties of the filamentary structure in them.

Giant molecular cloud G174+2.5 is situated in the Perseus Spiral Arm. Massive stars formed in the cloud belong to the Auriga OB2 association. Results of $^{12}$CO and $^{13}$CO observations of the cloud by~\citet{Heyer_1996} show complex kinematic structure of the gas. They find evidence for the compression of the molecular gas by \hii\ regions and mention interconnected molecular filaments in the vicinity of optical \hii\ regions. A list of the \hii\ regions as well as the their mutual location and location of young stellar clusters can be found in recent studies by \citet{Bieging_2016} and \citet{Ladeyschikov_2016}. \citet{Kirsanova_2008, Kirsanova_2014} and \citet{Dewangan_2016} present evidence of triggered star formation process in the G174+2.5 cloud around the brightest \hii\ region Sh2-235 (S235 below), Sharpless catalogue~\citet{Sharpless_1959}. \citet{Ladeyschikov_2015} also propose triggered star formation near \hii\ region Sh2-233 (S233 below). \citet{Ladeyschikov_2016} find a large filamentary molecular structure with dense clumps and embedded young stellar clusters on the border of G174+2.5 to the west from the extended \hii\ region Sh2-231 (S231 below): G173.57+2.43 associated with IRAS~05361+3539, S233\,IR asociated with IRAS~05358+3543, WB89-673 (WB\,673 below) and WB89-668 (WB\,668 below) associated with IRAS~05345+3556 and IRAS~05335+3609, respectively, see~\citet{Wouterloot_1989}. The center of the filamentary structure is given by WB\,673. Analysis of $^{12}$CO and \co\ emission shows that the clumps are gravitationally unstable,~\citet{Ladeyschikov_2016}. Their masses range from about 1000 to 2000\,M$_{\odot}$.

\begin{figure}
\includegraphics[width=8.5cm]{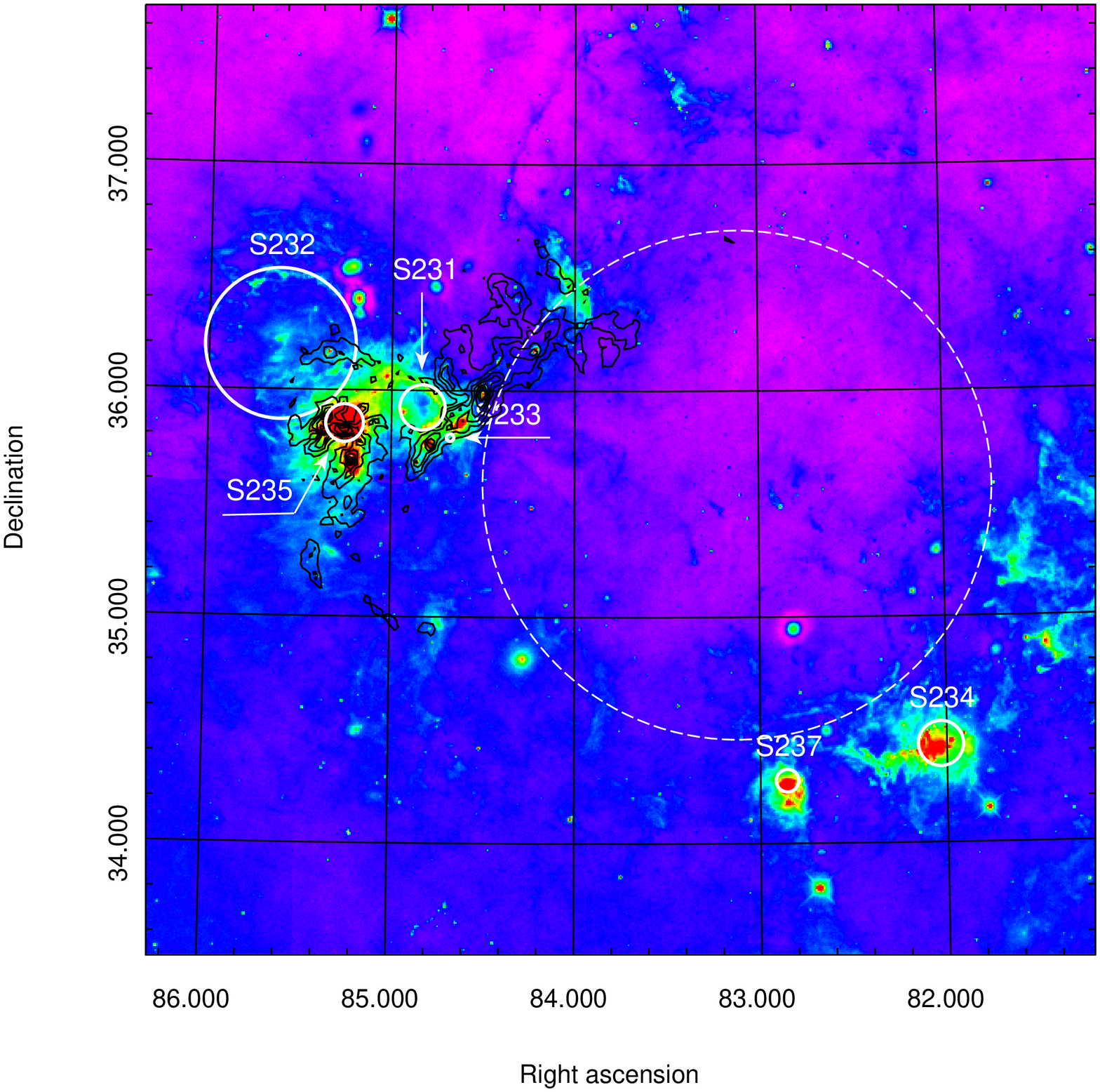}
\caption{Image at 22$\mu$m by WISE centered at $\alpha({\rm J2000.0}) = 05^{\rm h}35^{\rm m} 00^{\rm s}$ and $\delta({\rm J2000.0}) = +35^{\circ}36{'}36{''}$. The intensity scale uses an asinh stretch. Black contours represent $^{13}$CO emission of giant molecular cloud G174+2.5 from FCRAO\,14-m telescope by \citet{Ladeyschikov_2016}. Contours are generated from 1 to 150\,K~\kms. White circles show optical \hii\ regions from the Sharpless catalogue~\cite{Sharpless_1959}. White dashed circle shows unidentified bubble-shaped nebula.}
\label{fig:wise}
\end{figure}

Inspection of infrared images by the Wide-field Infrared Survey Explorer (WISE) and Herschel reveals a large unidentified bubble-shaped nebula in between S231--S235 star-forming region and \hii\ regions Sh2-234 and Sh2-237. Image of the nebula at 22$\mu$m by WISE is shown in Fig.~\ref{fig:wise}. The large filamentary molecular structure from \citet{Ladeyschikov_2016} is observed toward the place of possible intersection of the nebula and \hii\ regions S231 and Sh2-232. \citet{Kang_2012} discuss an old supernova remnant FVW~172.8+1.5 whose center and south-west arc coincide with the white dashed circle in our Fig.~\ref{fig:wise}. \citet{Jose_1017} also found this envelope on a Herschel-SPIRE image and study a star-formation in the south-western arc toward Sh2-234. So the filamentary structure could be an interesting example to test theoretical models by e.g.~\citet{Inutsuka_2015}.

The aim of our study is to look for a dense gas in the inter-clump medium of the filamentary structure and check if it is a real physically connected structure or the clumps are separated from each other but embedded into a common gas envelope. We call the whole filamentary structure WB\,673 below.

\section{Observations}\label{sec:obs}

We mapped emission in the \cs\ and \co\ transitions towards WB\,673 with the Onsala Space Observatory 20-m telescope in December 2016 and February 2017. We used a 3\,mm dual polarization sideband separating receiver by~\citet{Belitsky_2015} with an FFTS in 2$\times$2.5\,GHz mode. The LSB was tuned to 97.2\,GHz and USB to 109.2\,GHz. The spectral resolution was 76\,kHz. The observations were done in frequency-switch mode with the frequency offset 5\,MHz. The main lines of our interest were \co\ at 110.20\,GHz in USB and \cs\ at 97.98\,GHz in LSB. We also got \cvo\ at 109.78\,GHz, C$^{34}$S(2-1) at 96.41\,GHz and methanol series at 96.7\,GHz simultaneously with the main lines. Typical system temperature was from 80 to 250\,K for LSB and from 160 to 340\,K for USB. We excluded data with the system temperature higher than 500\,K from the analysis. Typical RMS level for our data is 0.3\,K in $T_{\rm mb}$ scale for \cs\ and 1.1\,K for \co. We made full sampling observations with $20{''}$ step towards the dense clumps and the inner part of the filament. The outer parts were mapped with $40{''}$ sampling. Data were reduced with the GILDAS\footnote{http://www.iram.fr/IRAMFR/GILDAS} software. Line profile fitting of \cs, \co\ and \cvo\, was done by gauss function using {\it curve\_fit} from {\it python} {\it scipy.optimize}. All analysis of the methanol lines was done with GILDAS.

\section{Spatial distribution of the molecular gas}\label{sec:spatial}

\begin{figure}
\includegraphics[width=8.3cm]{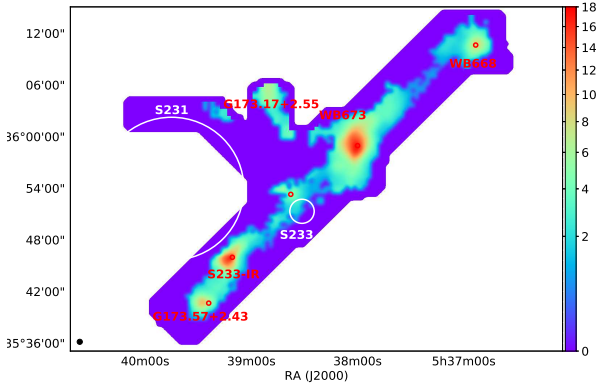}
\caption{\cs\ integrated emission in the WB\,673 filament. Colorbar shows intensity scale in K~\kms. White circles outline S231 and S233 \hii\ regions from the Sharpless catalogue~\cite{Sharpless_1959}. Red circles show positions where analysis of the methanol emission was done, see Sect.~\ref{sec:methanol}. Pixels with signal-to-noise ratio higher than 3 are shown on the map. Beam size $38.5''$ is in the left low corner. Bicubic interpolation is used to produce the map. The intensity scale uses a square-root stretch. \label{fig:maps}}
\end{figure}

The main result of the mapping is the detection of the \cs\ emission in the inter-clump medium. The dense clumps G173.57+2.43, S233\,IR, WB\,673 and WB\,668 are not isolated from each other. They are connected into a large filament. Fig.~\ref{fig:maps} shows a \cs\ emission map. Only pixels with signal-to-noise ratio higher than 3 are shown in Fig.~\ref{fig:maps}. We do not show a map of the \co\ emission here because there are recent already published larger maps by \citet{Bieging_2016}. The emission in \cs, \co\ and \cvo\ lines is detected from the south-east to the north-west of the filament. The emission peaks in \co\, \cvo\ and \cs\ lines are given by the WB\,673 dense clump. Brightness of \cs\ in S233\,IR is almost the same as in WB\,673. Emission in C$^{34}$S(2-1) line is not detected with signal-to-noise ratio higher than 3. We assume the distance to the filament to be $d=1.8$\,kpc after \citet{Evans_1981} and zero projection angle and determine that the length of the filament from WB\,668 in the north-west to G173.57+2.43 in the south-east to be 24.9\,pc. So the WB\,673 filament is four and two times longer than the integral shaped filament in the Orion\,A, see \citet{Hacar_2017} and NGC~\,6334 filament, \citet{Zernickel_2013}, respectively.

We calculate the column density of CS molecules \ncs\ in LTE using standard approach described by \citet{Mangum_2015}, Eq.80. In the beginning \ncs\ is calculated under assumption that \cs\ line is optically thin but then optical depth correction factor, \citet{Goldsmith_1999}, is used to make the value of \ncs\ reliable. For diatomic linear molecules: \[ N^{\rm thin }=\frac{3h}{8 \pi^3 S \mu^2} \times \frac{Q_{\rm rot}}{g_J g_K g_I} \times \frac{{\rm exp}\left(\frac{E_{\rm u}}{T_{\rm ex}}\right)}{{\rm exp}\left(\frac{h\nu}{kT_{\rm ex}}\right)-1 } \] \[  \times \frac{1}{(J_{\nu}(T_{ex})-J_{\nu}(T_{bg}))}  \int \frac{T_R dv}{f},\] where total molecular partition function $Q_{\rm rot} \simeq \frac{kT_{ex}}{hB}+\frac{1}{3}$, $g_J=2J+1, g_K=1, g_I=1$, $S=\frac{J^2}{J(2J+1)}$, $J_{\nu}(T)=\left( \frac{h\nu}{k}\right)/ \left( {\rm exp} \left( \frac{h\nu}{kT} \right) -1 \right)$. We assume filling factor $f=1$, $T_{\rm bg}=2.7$\,K and $\int T_{\rm R} dv$ defined as integrated intensity on $T_{\rm mb}$ scale. Other constants we use are given in Table~\ref{tab:micro}. Excitation temperature $T_{\rm ex}=20$\,K for \cs\ in LTE is taken the same as for CO lines in the same region by \citet{Bieging_2016}. To calculate the optical depth correction factor $\tau / (1-{\rm exp}(-\tau))$, we use the ratio of C$^{34}$S(2-1) to \cs\ line intensities: \[ \frac{T_{\rm mb}({\rm C^{34}S(2-1)})}{T_{\rm mb}({\rm CS(2-1)})} = \frac{1-{\rm exp}(-\tau_{\rm CS(2-1)} / r) }{1-{\rm exp}(-\tau_{\rm CS(2-1)}) }, \] e.g. \citet{Zinchenko_1994}. We use relative abundance ratio of CS to C$^{34}$S of $r=22.5$,~\citet{Wilson_1999}. This equation was solved using {\it fsolve} from {\it python} module {\it scipy.optimize}.

\begin{table}
\caption{Constants for \cs\ and \co\ column density calculations from NIST}
\begin{tabular}{lll}
\hline
              & \cs                  & \co                \\
\hline
$\mu$ (esu)   & $1.96\cdot 10^{-18}$ & $1.1\cdot 10^{-19}$\\
$B_0$ (MHz)   & 24495.576            & 55101.014          \\
$E_{u}$ (K)   & 7.1                  & 5.3                \\
$J_u$         & 2                    & 1                  \\
\hline
\end{tabular}
\label{tab:micro}
\end{table}

\ncs\ towards the emission peak in WB\,673 is up to $8\times10^{14}$\,cm$^{-2}$. Mean value of $\tau_{\rm CS(2-1)}$ is 5.7. Adopting relative abundance of CS to H$_2$ molecules $x_{\rm CS}$  in the range $10^{-9}-10^{-8}$, see e.g.~\citet{vanDishoeck_1998}, we determine the total mass of the dense gas in the filament: \[ M_{\rm dense} =  a_{\rm pix} \times 2.8 m_{\rm H} \times x_{\rm CS}^{-1} \times \sum_{\rm pix} N_{\rm CS},\] where $a_{\rm pix}$ is the pixel area and 2.8 is the mean molecular weight. This calculation gives a range $80<M_{\rm dense}<800$\,M$_{\odot}$ and mass-to-length ratio $3.4-34$\,M$_{\odot}$\,pc$^{-1}$ for the dense gas. We estimate the hydrogen mass surface density $\Sigma_{\rm H}$ of the dense gas as 700-7000\,M$_{\odot}$\,pc$^{-2}$ towards WB\,673 and up to 2700--27000\,M$_{\odot}$\,pc$^{-2}$ towards S233\,IR.

We do the same analysis in LTE for \co\ and \cvo\ lines to determine the CO column density $N_{\rm CO}$ and compare our results with \citet{Bieging_2016}, who did it for $^{12}$CO and $^{13}$CO. Isotopic ratio for our analysis is $r=8$ for $^{13}$CO/C$^{18}$O, e.g.~\citet{Wilson_1999}, and $r=80$ for $^{12}$CO/C$^{13}$O,~\citet{Bieging_2016}. We assume $T_{\rm ex}=20$\,K again. Mean value of $\tau_{\rm ^{13}CO(1-0)}$ is 5.0. $N_{\rm CO} \approx 10^{18}$\,cm$^{-2}$ towards the emission peak in WB\,673. Using CO relative abundance $x_{\rm CO}=10^{-4}$, see e.g.~\citet{vanDishoeck_1998}, we get peak $\Sigma_{\rm H}$ value of 17000\,M$_{\odot}$\,pc$^{-2}$ toward WB\,673 and 12000\,M$_{\odot}$\,pc$^{-2}$ to Sh2-233 and S233\,IR. This result is in agreement with LTE analysis by \citet{Bieging_2016}. The total mass of the gas in WB\,673 filament is $10^4$\,M$_{\odot}$, mass-to-length ratio is 360\,M$_{\odot}$\,pc$^{-1}$. Mass, length and mass-to-length ratio for the filament are comparable with other filaments identified in the Northen sky by \citet{Wang_2016}.

\section{Physical conditions in the dense clumps}\label{sec:methanol}

\begin{figure}
\includegraphics[width=7cm]{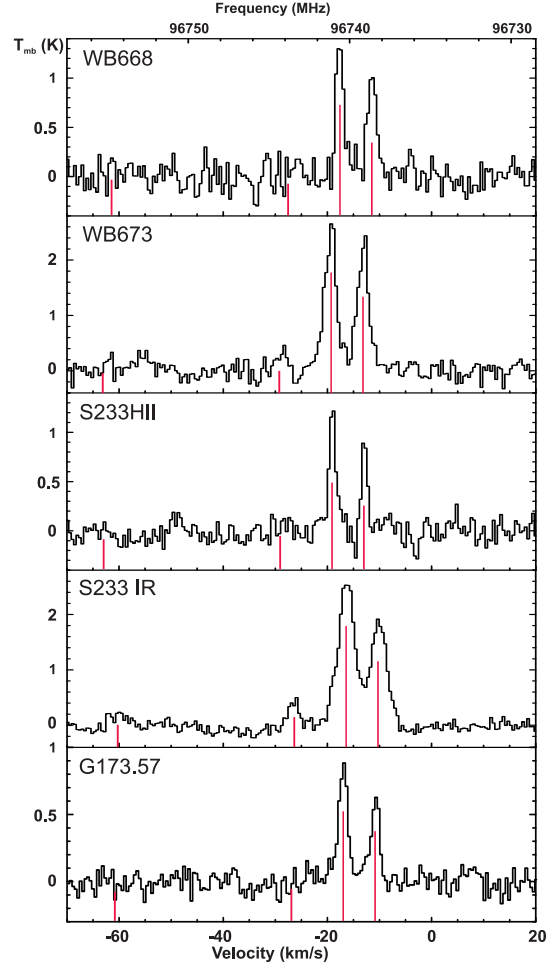}
\caption{Methanol emission lines toward the dense clump. Positions of the detected and non-detected methanol lines are shown by red lines.
\label{fig:meth_specs_at_pos}}
\end{figure}

Methanol emission is a diagnostic tool to estimate physical parameters of molecular clouds, e.g. \citet{Salii_2006, Salii_2006_ii, Leurini_2007, Zinchenko_2015}. All five clumps in the WB\,673 filament are detected using a quartet of quasi-thermal (i.e., not maser) methanol lines. Fig.~\ref{fig:meth_specs_at_pos} shows the methanol lines in the clumps. The brightest methanol lines $2_{-1}-1_{-1}\,v_t=0\,E$ and $2_{0}-1_{0}\,v_t=0\,A^{++}$ (96.739 \& 96.741~GHz) are confidently (above $5\,\sigma$) detected in all clumps, $2_{0}-1_{0}\,v_t=0\,E$ lines at 96.744~GHz are detected in 3 clumps with level about or above $3\,\sigma$, and $2_{1}-1_{1}\,v_t=0\,E$ emission at 96.755~GHz is detected in only one with level about $2\,\sigma$ (Table.~\ref{tab:meth_obs_positions}).

\begin{table}
\caption{Methanol lines quartet at 96.7~GHz has been detected in the clamps. Offset positions are given relative to $\alpha({\rm J2000.0}) = 05^{\rm h}38^{\rm m} 00^{\rm s}$ and $\delta({\rm J2000.0}) = +35^{\circ}59{'}17{''} 00$.}
\begin{tabular}{cccc}
\hline
 Notation   & $\int T_{br}dV$& $\Delta V$& $T_{\rm mb}$ \\
            & K*km/s         & km/s      & K            \\
\hline
\multicolumn{4}{l}{WB\,668 (-820$''$,680$''$)  $V_{LSR} = -17.60\pm 0.05 $~km/s } \\
{\small $2_{-1}-1_{-1}\,v_t=0\,E$}    & $2.3\pm 0.2$ &	$2.1\pm 0.2$&	$1.0\pm 0.2$\\
{\small $2_{0}-1_{0}\,v_t=0\,A^{++}$}  & $2.6\pm 0.2$ &	$1.7\pm 0.1$&	$1.4\pm 0.2$\\
{\small $2_{0}-1_{0}\,v_t=0\,E$}       & $0.4\pm 0.3$ &	            &   $\lesssim 0.2$ \\
{\small $2_{1}-1_{1}\,v_t=0\,E$}       & $0.2\pm 0.1$ &	            &   $\lesssim 0.2$\\
\multicolumn{4}{l}{WB\,673	(0$''$,-20$''$)	$V_{LSR} = -19.31\pm 0.03 $~km/s}\\	 							
{\small $2_{-1}-1_{-1}\,v_t=0\,E$}    & $6.3\pm 0.2$ &	 $2.7\pm 0.1$&	 $2.2\pm 0.2$\\
{\small $2_{0}-1_{0}\,v_t=0\,A^{++}$}  & $7.3\pm 0.2$ &	 $2.8\pm 0.1$&	 $2.5\pm 0.2$\\
{\small $2_{0}-1_{0}\,v_t=0\,E$}       & $0.9\pm 0.2$ &	 $2.8\pm 0.6$&	 $0.3\pm 0.2$\\
{\small $2_{1}-1_{1}\,v_t=0\,E$}       & $1.5\pm 0.7$ &	             &	 $\lesssim 0.2$ 	\\						
\multicolumn{4}{l}{Sh2-233	 (460$''$,-360$''$)	$V_{LSR} = -19.00\pm 0.03 $~km/s }\\									 		
{\small $2_{-1}-1_{-1}\,v_t=0\,E$}    & $1.3\pm 0.1$ &	$1.2\pm 0.1$&	$1.0\pm 0.2$\\
{\small $2_{0}-1_{0}\,v_t=0\,A^{++}$}  & $2.1\pm 0.1$ &	$1.6\pm 0.1$&	$1.2\pm 0.2$\\
{\small $2_{0}-1_{0}\,v_t=0\,E$}       & $0.3\pm 0.2$ &	            &	$\lesssim 0.2$\\
{\small $2_{1}-1_{1}\,v_t=0\,E$}       & $0.1\pm 0.1$ &	            &   $\lesssim 0.2$\\
\multicolumn{4}{l}{S233\,IR	(860$''$,-800$''$)	$V_{LSR} = -16.27\pm 0.03 $~km/s}\\				
{\small $2_{-1}-1_{-1}\,v_t=0\,E$}    & $7.3\pm 0.2$ &	$3.6\pm 0.1$&	$1.9\pm 0.1$\\
{\small $2_{0}-1_{0}\,v_t=0\,A^{++}$}  & $10.5\pm 0.2$ &	$3.8\pm 0.1$&	$2.6\pm 0.1$\\
{\small $2_{0}-1_{0}\,v_t=0\,E$}       & $1.1\pm 0.1$ &	$2.3\pm 0.3$&	$0.5\pm 0.1$\\
{\small $2_{1}-1_{1}\,v_t=0\,E$}       & $0.8\pm 0.1$ &	$3.6\pm 0.6$&	$0.2\pm 0.1$\\
\multicolumn{4}{l}{G173.57+2.43 (1060$''$,-1160$''$)	$V_{LSR} = -16.27\pm 0.03 $~km/s}\\	
{\small $2_{-1}-1_{-1}\,v_t=0\,E$}    & $1.0\pm 0.1$ &	$1.5\pm 0.2$&	$0.7\pm 0.2$\\
{\small $2_{0}-1_{0}\,v_t=0\,A^{++}$}  & $1.7\pm 0.1$ &	$1.4\pm 0.1$&	$1.1\pm 0.2$\\
{\small $2_{0}-1_{0}\,v_t=0\,E$}       & $0.2\pm 0.1$ &	        &    $\lesssim 0.2$\\
{\small $2_{1}-1_{1}\,v_t=0\,E$}       & $0.3\pm 0.7$ &	        &    $\lesssim 0.2$\\
\hline 
\end{tabular}
\label{tab:meth_obs_positions}
\end{table}

Shift of radial velocities of the methanol lines is observed between the clumps, see Fig.~\ref{fig:meth_specs_at_pos} and Table.~\ref{tab:meth_obs_positions}. The velocity changes from $-19.3$\,\kms\ in the central position of WB\,673 clump, to $-17.6$\,\kms\ in the north-west towards WB\,668, and up to $-16.3$\,\kms\ in the south-east towards G173.57+2.43. A similar velocity shift is observed in \co\ and \cs\ lines, see Sect.~\ref{sec:vel}. 

To estimate the physical conditions in the regions emitting methanol lines, we use a simple radiative transfer model, which uses the LVG approximation. Our model has four parameters: gas kinetic temperature (\tk), hydrogen number density ($n_{\rm H_2}$,~cm$^{-3}$), methanol specific column density (\nmeth, column density per unit velocity) and methanol relative abundance ($x_{\rm CH_3OH}=N_{\rm CH_3OH}/N_{\rm H_2}$). Dust emission and absorption within the emission region is taken into account in the way described in \citet{Sutton_2004}. Since it is not known exactly if the cloud fill the beam totally or not, a filling factor, $f=0.95$, is included to the line intensity calculation. The scheme of energy levels in this model includes rotational levels with quantum numbers J up to~22 and |K| up to 9; the levels include the rotational levels of the ground, first and second torsionally excited states. In total, 861 levels of A-methanol and 852 levels of E-methanol were considered according to \citet{Cragg_2005}.
 
In order to estimate the physical parameters, we look for a set of parameters that exhibits the best agreement between the values of the calculated brightness temperatures $T_i^{mod}$ and the measured brightness temperatures $T_i^{obs}$. This corresponds to finding the minimum of \[\chi^2 = \frac{1}{N} \times \sum_i^N{\left(\frac{ T_i^{obs} - T_i^{mod}}{\sigma_i}\right)^2},\] where $\sigma_i$ is the observational uncertainty for a particular line and $N$ is the total number of explored lines .

We explore the parameter space to find the approximate location of the $\chi^2$ minimum. For this purpose, we use the database of population numbers for the quantum energy levels of methanol by \citet{Salii_2006}. The \tk\ in the database ranges from 10 to 220~K, the $n_{\rm H_2}$~--- from $10^3$ to $10^9$~cm$^{-3}$, the \nmeth--- from $10^{8}$ to $10^{13}$~cm$^{-3}$s, the relative abundance of methanol molecules relative to molecular hydrogen  ~--- from $10^{-9}$ to $10^{-6}$, \citet{vanDishoeck_1998,Sutton_2004,Zinchenko_2015}. The lower values of these parameters correspond to the physical conditions of dark molecular clouds, while the higher values of both parameters can occur in shocked molecular material. Thus we can use this base to investigate molecular clouds at their different stages.

The set of the methanol lines is not very sensitive to variations of the $x_{\rm CH_3OH}$ for the dense clumps under consideration. Taking into account hydrogen column density estimations by \co\ and \cs\ lines in Sect.~\ref{sec:spatial} we find a reliable value of the $x_{\rm CH_3OH}=10^{-8}-10^{-7}$. With such limitat we estimate \tk$=15-35$~K, \nmeth$=1.8\times10^{9}$ and $10^{9}$~cm$^{-3}$s for WB\,673 and  S233\,IR, respectively, and $n_{\rm H_2}=10^4$~cm$^{-3}$ for both sources. The \tk\ values are in good agreement with the results of~\citet{Bieging_2016}.

Since there are few detected methanol lines at the other sources we use \tk\ from~\citet{Bieging_2016} and again $n_{\rm H_2}=10^4$~cm$^{-3}$ for all of them. In this way, \nmeth,\ are estimated as $5\times10^{8}$ and $3\times10^{8}$~cm$^{-3}$s for Sh2-233 and G173.57+2.43 respectively. Assuming \tk=20\,K for WB\,668, we get \nmeth=$5\times10^{8}$~cm$^{-3}$s too. We note that methanol abundance $x_{\rm CH_3OH}=10^{-8}-10^{-7}$ is greater than it is in dark clouds, so it can be explained by some star formation processes or by shock wave propagation.

\section{Velocity field}\label{sec:vel}

\begin{figure}
\includegraphics[width=8cm]{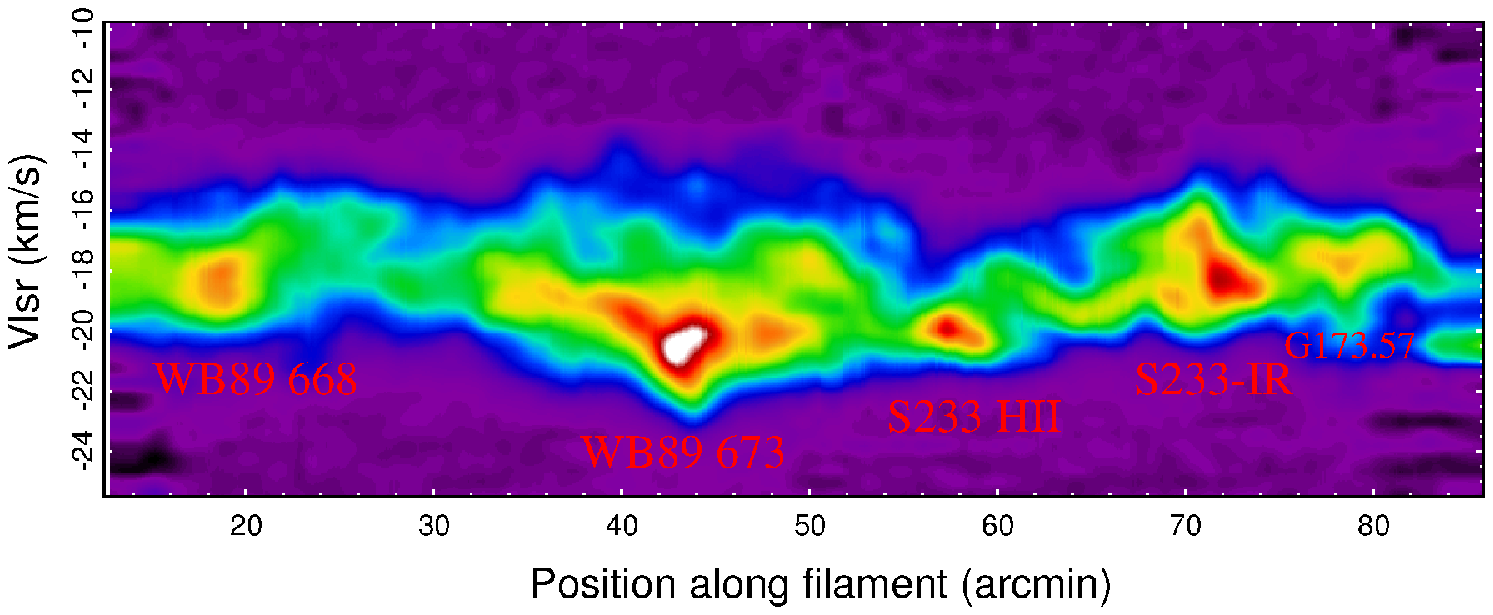}
\includegraphics[width=8cm]{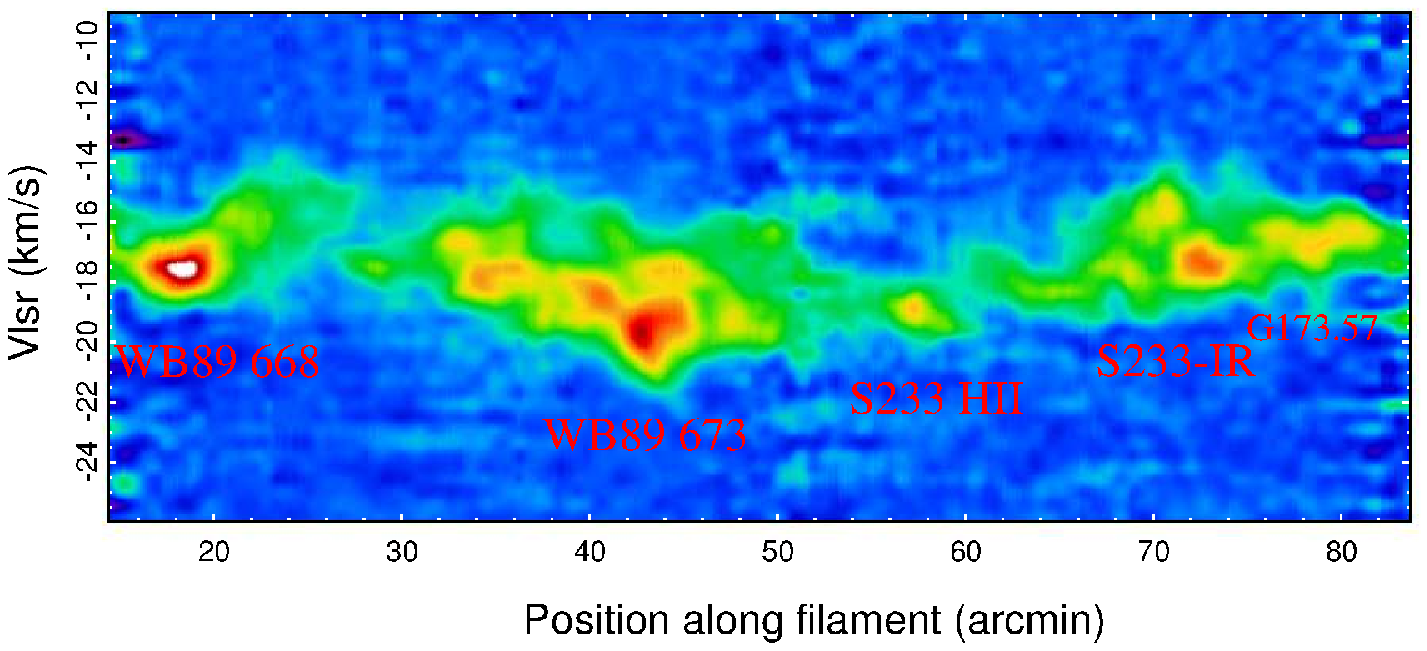}
\includegraphics[width=8cm]{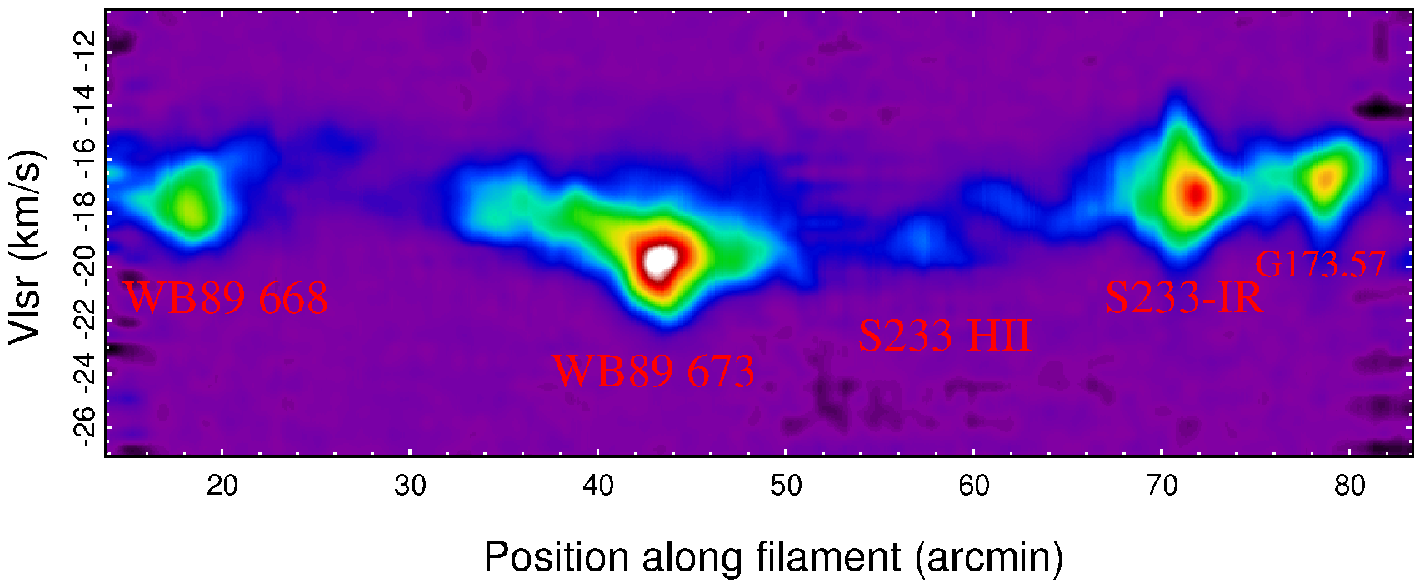}
\caption{Position-velocity diagrams for \co\ (top), \cvo\ (middle) and \cs\ lines (bottom).\label{fig:pv} Intensity in images increases linearly, with a maximum of 10\,K in \co, 1.5\,K in \cvo\ and 3.3\,K in \cs.}
\end{figure}

Position-velocity (PV) diagrams for the emission in \co, \cvo\ and \cs\ lines are shown in Fig.~\ref{fig:pv}. The PV-diagram are calculated along the filament from the north-west to the south-east using integration width of $360''$.  The diagrams show that radial velocities of the G173.57+2.43 and WB\,668, the south-eastern and the north-western ends of the filament, are red-shifted relative to the center at WB\,673 up to 2-3\kms. The shape of the PV-diagrams and values of the shift are almost the same for the three lines. We determine the gradient value about 0.2\kms\,pc$^{-1}$ from WB\,673 to G173.57+2.43 and from WB\,673 to WB\,668. A higher velocity gradient 0.6\,\kms\ is observed in between WB\,673 and the region without bright \cs\ emission toward $x=31'$ in Fig.~\ref{fig:pv}. 

\begin{figure}
\includegraphics[width=8cm]{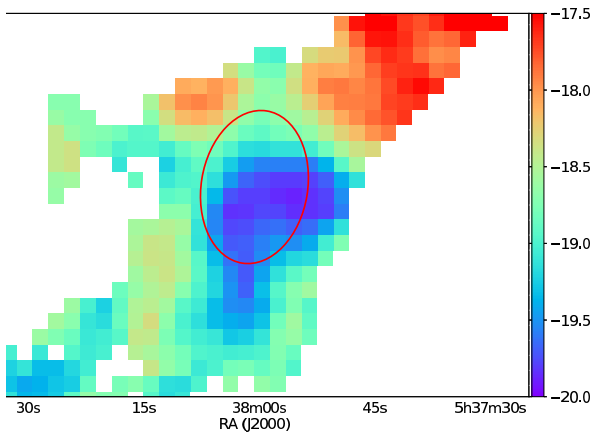}
\caption{Map of $V_{\rm lsr}$ distribution in WB\,673 dense clump from \cs\ data. Red ellipse shows size of WB\,673 from the fit of $^{12}$CO and $^{12}$CO data by~\citet{Ladeyschikov_2016}.}
\label{fig:csvel}
\end{figure}

Fig.~\ref{fig:csvel} shows the velocity distribution of \cs\ line in the center of the filament -- WB\,673 dense clump. The area with the most negative velocities coincides with the peak of \cs\ and methanol emission. Velocities at the periphery of the clump are more positive. The velocity difference of \cs\ lines between the center and the periphery of WB\,673 is 2\,\kms\ which is about a typical linewidth of \cs\ in WB\,673. The  difference is larger than the thermal linewidth for \tk=15-35\,K. The same velocity distribution was found by \citet{Kirsanova_2017} in star-forming clouds from the Perseus spiral arm G183.35-0.58 and also likely in G85.40-0.00. The value of the difference in WB\,673 is two times larger than in G183.35-0.58. Such velocity difference might be a result of the overflowing of molecular cloud by a large-scale wave, \citet{Kirsanova_2017}, but we avoid straightforward interpretation of the velocity field in WB\,673.

V-shaped PV-diagram could be an indication of gravitational collapse of the ends to the center of the filament. Red-shifted ends are observed in the integral shaped filament in the Orion\,A cloud by \citet{Hacar_2017}. They applied a model of a free-fall of point-like mass to the center, considering that the mass accumulated in the center of the filament is about twice the mass in the ends. Blue-shifted ends relative to the center are observed in NGC\,6334 filament by~\citet{Zernickel_2013}. They also mention that the V-shaped PV-diagram could be an attribute of the gravitational collapse. 

We do not apply other analysis such as point-like mass in free-fall to WB\,673 filament because the mass distribution and absolute value of the velocity gradient in WB\,673 are different. \citet{Ladeyschikov_2016} estimate masses of the dense clumps in the WB\,673 filament using $^{12}$CO and $^{13}$CO data. They found WB\,673 to be two times more massive than WB\,668, their masses are 2100 and 1200\,M$_{\odot}$, respectively. So, the mass ratio of the central to the outermost clumps is the same as in the Orion\,A filament. Total mass of S233\,IR and G173.57+2.43 is 1980\,M$_{\odot}$, that is almost same as WB\,673. We note that the absolute values of the velocity gradients in the Orion\,A and NGC\,6334 are 5-10 times larger than in the WB\,673 filament. But masses of the clumps are comparable to those detected by \citet{Hacar_2017} in Orion~A. So their numerical values, shown in their Fig.\,2, are not appropriate for the velocity gradient  in the WB\,673 filament. We postpone the analysis of the gas kinematics in details for the next study.


\section{Conclusion}\label{sec:conc}

The high-mass filament WB\,673 in G174+2.5 giant molecular cloud could be a good example of a filament whose formation was influenced by expanding shells: \hii\ regions S231 and S232 from one side and from the other side the unidentified shell-like nebula visible on infrared images by WISE and Herschel. 

We find dense gas in WB\,673 filament not only toward clumps with embedded stellar clusters but also in the inter-clump medium. So the filament is a large connected structure with the total mass $10^4$\,M$_{\odot}$ and mass-to-length ratio 360\,M$_{\odot}$\,pc$^{-1}$. Mass-to-length ratio for the dense gas is $3.4-34$\,M$_{\odot}$\,pc$^{-1}$ from \cs\ data. These parameters of the filament are comparable with other filaments identified in the Northen sky by \citet{Wang_2016}. V-shaped PV-diagram of the filament has red-shifted ends relative to the center. This shape could be a signature of gravitational contraction of the filament. Our analysis of the methanol emission in the dense clumps gives temperatures and densities in agreement with the results from $^{12}$CO and $^{13}$CO analysis by \citet{Bieging_2016}. The abundance of methanol is higher than in cold dark clumps. It can be explained by some star formation processes or by shock wave propagation from the embedded stellar clusters.

We conclude that the WB\,673 filament is a promising region to study feedback from 'older' generation of stars to forming 'younger' generation.

\section*{Acknowledgments}
We are thankful to S. Yu. Parfenov and D. S. Wiebe for useful discussions.

MSK is supported by Program of Fundamental Research of the Presidium of the RAS and President of the Russian Federation grant NSh-9576.2016.2. SVS is supported by Act 211 Government of the Russian Federation, contract № 02.A03.21.0006.


\end{document}